%% file: main.tex
\documentclass[conference]{IEEEtran}
\IEEEoverridecommandlockouts
\usepackage{cite}
\usepackage{enumitem}
\usepackage{amsmath,amssymb,amsfonts}
\usepackage{algorithmic}
\usepackage{graphicx}
\usepackage{textcomp}
\usepackage{caption}
\usepackage{makecell}
\usepackage{xcolor}
\def\BibTeX{{\rm B\kern-.05em{\sc i\kern-.025em b}\kern-.08em
    T\kern-.1667em\lower.7ex\hbox{E}\kern-.125emX}}
\usepackage{hyperref}
\begin{document}
\title{Quantization-aware Neural Architectural \\Search for Intrusion Detection
}

\author{\IEEEauthorblockN{Rabin Yu Acharya*, Laurens Le Jeune$\dagger$, Nele Mentens$\dagger$, Fatemeh Ganji$\ddagger$, Domenic Forte*}
\IEEEauthorblockA{{*University of Florida, $\dagger$ KU Leuven, $\ddagger$ Worcester Polytechnic Institute}\\
rabin.acharya@ufl.edu, \{laurens.lejeune,nele.mentens\}@kuleuven.be, fganji@wpi.edu, dforte@ece.ufl.edu}
\and
}

\maketitle

\begin{abstract}
Deploying machine learning-based intrusion detection systems~(IDSs) on hardware devices is challenging due to their limited computational resources, power consumption, and network connectivity. Hence, there is a significant need for robust, deep learning models specifically designed with such constraints in mind. In this paper, we present a design methodology that automatically trains and evolves quantized neural network (NN) models that are a thousand times smaller than state-of-the-art NNs but can efficiently analyze network data for intrusion at high accuracy. 
In this regard, the number of LUTs utilized by this network when deployed to an FPGA is between $2.3\times$~and~$8.5\times$ smaller with performance comparable to prior work.
\end{abstract}


\input{Introduction}
\input{related_work}
\input{method}
\input{results}
\input{conclusion} 
\section{Acknowledgments}
This work has been supported partially by NSF under award numbers 2117349 and 2138420.  

\bibliographystyle{ieeetr}
\bibliography{references}

\end{document}

%% file: Introduction.tex
\section{Introduction}\label{sec:intro}
In today's interconnected world, where data breaches and cyberattacks pose important risks to individuals and organizations, safeguarding computer networks and systems has become paramount. One crucial tool in the arsenal of cybersecurity professionals is the Intrusion Detection System (IDS). An IDS is a security technology that monitors and analyzes network traffic or host activities, aiming to detect and respond to unauthorized or malicious activities. By doing so, IDSs play a vital role in identifying potential threats and enabling proactive defense measures~\cite{khraisat2019survey}. 
However, as network traffic bandwidth continues to grow, there is a greater demand for network infrastructure to achieve higher throughput. Thus, IDSs need to effectively identify attacks within these high-traffic environments.

In recent years, machine learning techniques have significantly enhanced the capabilities of IDSs. Traditional IDSs relied on rule-based approaches, where predefined rules were used to identify known patterns of attacks~\cite{debar1999towards}. While explainable, these rule-based systems struggle to keep up with the ever-evolving nature of cyber threats. Machine learning techniques, on the other hand, have the ability to learn from large volumes of data and identify complex patterns~\cite{9410557}. These algorithms can analyze various features of network traffic, such as packet headers, traffic volumes, protocols, and timing information, to identify anomalies that may indicate a potential intrusion. However, deploying machine learning-based IDSs on hardware or edge devices poses several challenges. Edge devices, such as routers, Internet-of-Things (IoT) devices, or embedded systems, typically have limited computational resources, power constraints, and restricted network connectivity. These barriers make it difficult to implement complex machine-learning algorithms directly on the edge. A major challenge is the high computational requirements of machine learning models, which often involve complex computations and large parameter spaces. Edge devices do not have sufficient processing power or memory capacity to handle these resource-intensive tasks. Additionally, machine learning models typically require frequent updates and retraining to adapt to new threats, which can be challenging to achieve on edge devices with limited connectivity.

Despite these challenges, efforts are being made to address the deployment issues of machine learning-based IDSs at the edge. One approach involves deploying lightweight machine-learning models specifically designed for edge devices. These models are optimized to operate efficiently with limited resources, using techniques like model compression~\cite{chen2014efficient}, quantization~\cite{murovivc2021genetically, umuroglu2020logicnets, vrevca2021detecting, le2022feature}, and knowledge distillation~\cite{wang2022lightweight}. By reducing the model complexity and memory footprint, these lightweight models can be deployed on edge devices while maintaining acceptable intrusion detection performance.

In this paper, we provide a new methodology to design compact yet deep-enough neural networks that use significantly fewer resources compared to state-of-the-art machine learning systems. More specifically our contributions are as follows:
\begin{enumerate}[leftmargin=*,nosep]
    \item We employ the neural architectural search methodology to design compact neural networks to perform highly accurate and robust classification on one of the most popular network-based datasets, the UNSW-NB15~\cite{moustafa2015unsw}. 
    \item We further showcase a methodology to perform quantization aware training~(QAT) to develop quantized neural networks which provide improvements via model compression and latency reduction to reduce the number of hardware resources required.
    \item We compare our performance against state-of-the-art techniques to demonstrate how we reduce the network size by more than a $1000\times$ and yet obtain comparable performance.
\end{enumerate}
The rest of the paper is organized as follows. Section~\ref{sec:related_works} provides background on related works. Section~\ref{sec:methodology} describes our proposed methodology and Section~\ref{sec:results} provides corresponding results while comparing them to other alternative techniques. Finally, Section~\ref{sec:conclusion} concludes the paper.

%% file: related_work.tex
\section{Related Work}\label{sec:related_works}
\subsection{Intrusion Detection System~(IDS)}
Cyberthreats come in various forms, ranging from traditional malware and viruses to sophisticated hacking techniques. Attackers exploit vulnerabilities in systems to gain unauthorized access, steal sensitive information, disrupt services, or compromise the integrity of data. Common types of threats include distributed denial of service~(DDoS) attacks, network scanning, intrusion attempts, malware infections, and insider threats. An IDS acts as a vigilant guardian, continuously scrutinizing network traffic or host behavior to identify indicators of these threats.
Traditionally, IDS systems comprised signature-based detection matching network or host activities against a database of known attack signatures, or expert systems defining specific rules~\cite{debar1999towards}. However, with the increasing complexity and polymorphism of modern threats, those traditional approaches alone are often insufficient, as they require frequent and manual updating and are unable to detect new attacks~\cite{gholami2021survey}. This is where deep learning systems have emerged as a promising avenue for enhancing IDS capabilities, with their potential to detect unknown attacks~\cite{Buczak2016}.

\subsection{Deep Learning-based IDS}
Deep learning, a subset of machine learning, leverages artificial neural networks (NNs) to automatically learn patterns and features from vast amounts of data. By training IDS models on massive datasets that include both benign and malicious network or host behavior, deep learning algorithms can identify intricate attack patterns and detect previously unknown threats.
The application of deep learning in IDS brings several advantages. It allows for the detection of complex attacks that may have evaded traditional signature-based systems, as deep learning models can identify subtle deviations from normal behavior. Additionally, deep learning-based IDS systems can adapt and evolve over time, improving their detection capabilities as they encounter new threats. 

Nevertheless, there are challenges in implementing deep learning-based IDS systems. One significant bottleneck is the need for large and diverse training datasets that accurately represent real-world attack scenarios. Acquiring and labeling such datasets can be time-consuming and resource-intensive. Moreover, deep learning models require substantial computational resources for training and inference, which can pose scalability issues for large-scale deployments. Thus, recent works have focused on deploying these systems on the edge or even alternative platforms such as Field Programmable Gate Arrays~(FPGAs). This enables computing systems to effectively use deep learning close to sensor devices without sending the data to the hosts, thus significantly reducing bandwidth requirements while achieving greater speed and better security. A binary neural network~(BNN) is proposed in~\cite{murovivc2021genetically, umuroglu2020logicnets, vrevca2021detecting} where the network is deployed on an FPGA. BNNs are essentially quantized state-of-the-art solutions that are built to use fixed point, integer or binary weights and activation functions, significantly reducing the memory footprint and computational requirements of the NN. Further, they can efficiently be implemented on an FGPA optimized for binary operations, enabling faster inference and lower power consumption.
Similarly, convolutional NNs (CNNs) have been designed for FPGA deployment by using feature reduction and processing techniques to improve the network's accuracy while optimizing for high throughput~\cite{le2022feature}.

In this paper, we use evolutionary techniques to evolve NNs from scratch such that the network's architecture and weights are optimized. Specifically, we adapt open-source code for this purpose called InfoNEAT~\cite{acharya2023information, Acharya_InfoNEAT}. Furthermore, we employ quantization-aware training techniques within InfoNEAT to evolve and train quantized NNs that can be efficiently deployed on an FPGA or edge device. 

%% file: method.tex
\begin{figure}[t]
\centerline{\includegraphics[width=0.4\textwidth,height=0.4\textwidth,keepaspectratio]{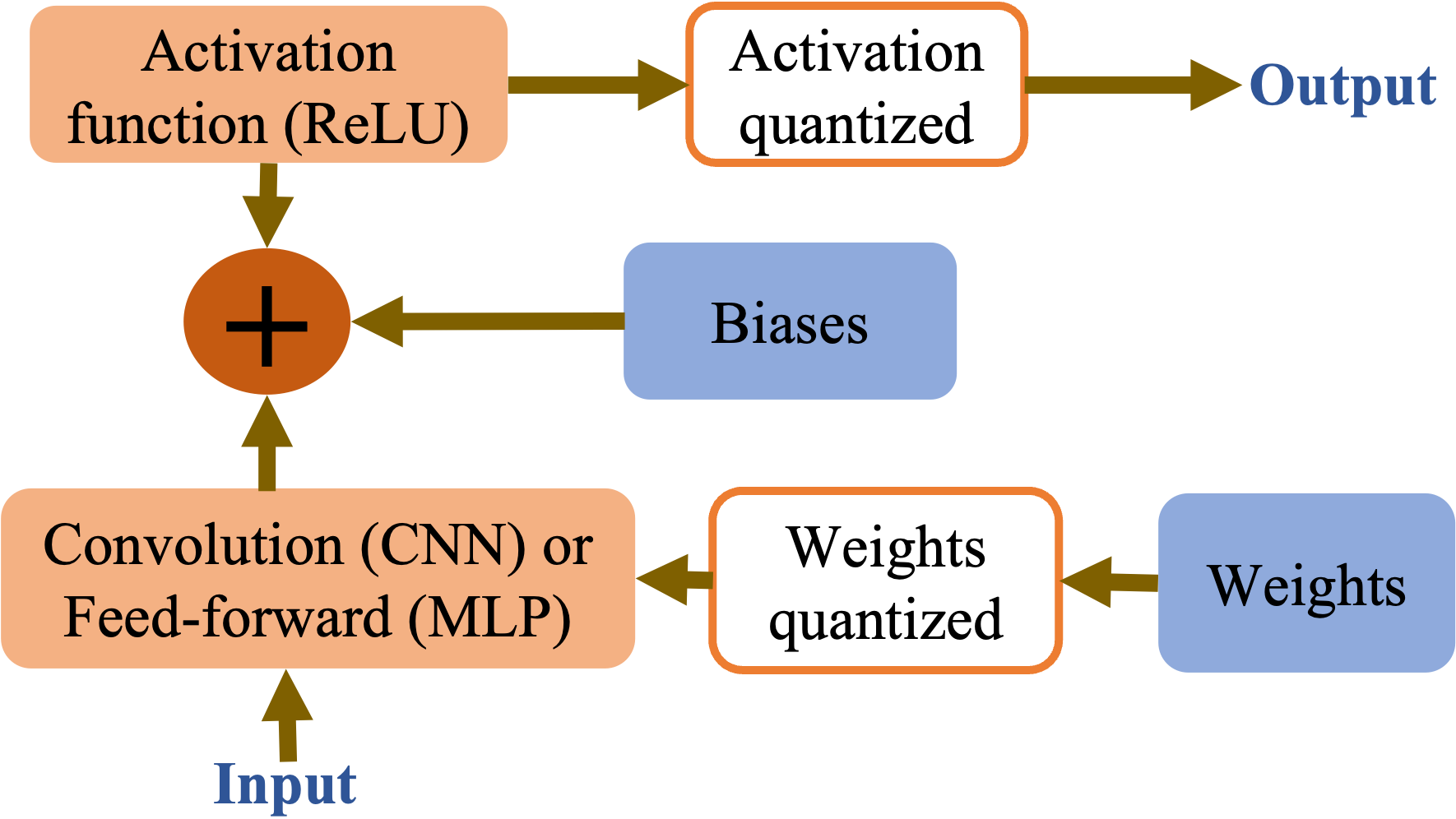}}
\caption[Overview of quantization aware training algorithm as used in q-InfoNEAT.]{Overview of quantization aware training algorithm.}
\label{fig:qat}
\end{figure}
\section{Design Methodology}\label{sec:methodology}
This section provides the design methodology used in this paper to design NNs for an IDS. Section~\ref{sec:infoneat} provides details on the InfoNEAT algorithm while Section~\ref{sec:method_qat} provides details on the quantization technique used to enhance InfoNEAT algorithm to be applied in an IDS. 
\subsection{InfoNEAT}\label{sec:infoneat}
Information Theoretic-based Neuroevolution of Augmenting Topologies (InfoNEAT) is an algorithm that revolves around the notion of the evolution of NNs or ``neuroevolution'' which evolves the architectures or configurations of multiple networks (so-called, augmenting topologies) and simultaneously tunes their hyperparameters. InfoNEAT has been introduced recently~\cite{acharya2023information, Acharya_InfoNEAT}, which was built upon NEAT (Neuroevolution of Augmenting Topologies)~\cite{stanley2002evolving}. 
The underlying evolutionary algorithm in NEAT optimizes hyperparameters of multi-layer perceptron (MLP)-like networks; specifically, the weights of individual neurons and their interconnections are evolved in each step, called a \textit{generation}. 
In fact, the configuration of NNs is different from typical MLPs, where each hidden neuron is connected to every neuron in the previous and next layers, as illustrated in Fig.~\ref{fig:dummy_mlp}. 
The irregular NN configuration (compared to the structure of MLPs) discovered by NEAT is one of its most important aspects and contributes to its special behavior. 

InfoNEAT applies information theoretic concepts to assist with the selection of the best NN and stop the evolution at the correct time. 
In doing so, the main disadvantage of selection method in NEAT have been addressed~\cite{stanley2002evolving}, i.e., saving time spent on selecting the topology manually. 
Here, we briefly describe the learning process in InfoNEAT and refer the interested reader to~\cite{acharya2023information}. 
The evolution process begins with a set of NNs with minimal complexity. Then these networks are evaluated based on a fitness function so that the algorithm can decide which NNs would be successful in the next generation. The selected individuals or networks are passed on to the next step, called \textit{Crossover}, which basically is a recombination step where the parameters of two different NNs are exchanged or combined. The rest of the individuals are ``mutated''. Specifically, random changes such as the removal of nodes, changes in connections, etc. are made to update their configurations and parameters. These steps -- selection, crossover, and mutation -- are repeated until a stopping condition is met. 

\vspace{10pt}
\noindent\textbf{Information-theoretic criteria. }
In NEAT, a maximum number of generations is considered to stop the training.
That is usually accompanied by a heuristic, e.g., if the accuracy of the best NN (i.e., the NN with the highest accuracy) does not improve after some generations, the evolution is halted. 
To improve the efficiency and avoid overfitting, InfoNEAT has applied information theory-based approaches to stop the evolution process. 
Intuitively, the evolution process is stopped as soon as the NNs can learn more, given the labeled inputs. 

Another issue with NEAT resolved in InfoNEAT is which NN should be chosen after the evolution is stopped. 
As said before, the NEAT's premise is to evolve NNs over many generations to obtain a large population of NNs including many \emph{species}, each with several \emph{genomes} (NNs). 
From one generation to the next, making an unnecessary change to the well-trained genome can lead to a reduction in how much the NN can learn from the labeled inputs~\cite{cover1999elements,vinh2014reconsidering}. 
The higher this decrease is, the less useful the change made to the genome would be. 
This has been translated to the amount of information extracted by the NNs to select the best NN thanks to the information theory-based metrics cf.~\cite{acharya2023information}.

\subsection{q-InfoNEAT}\label{sec:method_qat}
InfoNEAT has been tailored toward the needs of a machine learning tasks with multiple classes. 
Nevertheless, in a scenario, where an IDS aims to classify benign and malicious activities, a binary task should be tackled. 
Therefore, one of the major differences between the q-InfoNEAT and InfoNEAT proposed in~\cite{acharya2023information} is that q-InfoNEAT trains a single model. 
Hence the stacking method (that combines predictions from multiple trained models to create a meta-learner) as used in InfoNEAT is not used. Furthermore, and more importantly, to meet the needs for IDS running on a hardware platform, the model is quantized using the method described next.   

Alongside the InfoNEAT algorithm mentioned above, we integrate the quantization-aware training (QAT) framework introduced in~\cite{zhang2018lq} to train a quantized NN model. The so-called Learned Quantization or LQ-Nets framework combines learning-based quantization with network architecture optimization to achieve highly accurate and compact deep neural networks as shown in Fig.~\ref{fig:qat}. The key idea behind LQ-Nets is to jointly optimize the network architecture and the quantization levels of weights and activations during the training process. By integrating quantization into the learning process, LQ-Nets overcomes the limitations of traditional post-training quantization methods and achieves improved accuracy and compactness.
The work in~\cite{zhang2018lq} is mainly suited for CNN. However, we adapt their functions within InfoNEAT to evolve and train quantized MLP-like NNs.

The goal of any quantization method is to represent the floating-point weights or activations in as few bits as possible. Typically, a quantization function is a piecewise-constant function represented as:
\begin{equation}
    Q(x) = q_l, \text{if } x \in (t_l, t_{l+1}],
\end{equation}
where $q_l$ are the quantization levels and $(t_l, t_{l+1}]$ are the quantization intervals. Using this function, a floating point value is quantized to its nearest quantization level if the value falls within the quantization interval. During the training process, however, the weights and activations may change. So the problem becomes defining a quantizer that yields minimum quantization error (the squared sum difference between actual and quantized value). So rather than pre-defining a quantizer, the LQ-Net framework jointly trains the neural network as well as the quantizer. This not only minimizes the quantization error but also adapts well to the training goal thus improving the final performance or accuracy. 

Furthermore, the quantizers are trained in such a way that they are compatible with bitwise operations which is crucial to ensure efficient computation, storage, and inference on hardware~\cite{gholami2021survey}. For this purpose, the quantized values are represented in such a way that they can be decomposed by a linear combination of the bits. Specifically, the quantization function is represented as:
\begin{equation}
    Q(x,v) = q_l = v^{T}b_l, \text{if } x \in (t_l, t_{l+1}], l \in {1,..,2^k},
\end{equation}
where $v$~is the learnable basis vector, $b_l$ represents the corresponding binary encoding such that $b_l \in \{-1,1\}^k$ for weights and $b_l \in \{0,1\}^k$ for activation functions. Basis vector as the name suggests is a set of real numbers that form the basis for representing the quantized values where the number of elements in the vector corresponds to the number of quantization levels. 

To train LQ-Nets, we first use floating-point network weights and activation outputs to initialize a basis vector and calculate a corresponding binary encoding vector. This process is run for $T$~iterations to minimize the quantization error. The goal is to find optimal quantizer basis $v$~(or $B$) during training. This $v$~can also be represented as $(BB^T)^{-1}Bx$ where $B$~represents the set of $b_l$ for the input values.
These quantized weights and activations are then optimized using the InfoNEAT algorithm. After training, these weights can be discarded, however, their binary codes and quantizer bases are kept for testing~\cite{zhang2018lq}. 

\subsection{Deployment in Hardware}
After the model is trained, we then proceed to deploy this model on hardware. The training of optimized and quantized model using InfoNEAT in Sections~\ref{sec:infoneat} and~\ref{sec:method_qat} helps to significantly reduce memory requirements and computational complexity, making the model more suitable for hardware deployment. Regarding the hardware, we use FPGAs which are considered to be the preferable choice because of their performance, energy efficiency, reconfigurability, and scalability. 

The implementation of the NN model involves converting it into a hardware description language~(HDL) representation, which is used to program the FPGA. The HDL code describes the network's structure, connectivity, and computations in a format that the FPGA can understand. Once the network is implemented in HDL, hardware-specific optimizations are applied to further enhance the performance of the FPGA deployment. These optimizations leverage the parallelism and pipelining capabilities of FPGAs. Techniques such as loop unrolling, pipelining, and parallel processing are utilized to fully exploit the parallelism available on the FPGA and achieve higher throughput and efficiency. Memory hierarchy customization, efficient data movement schemes, and the use of specialized FPGA resources also play a significant role in optimization. Then, a communication interface is designed to interact with the FPGA-based NN. This interface facilitates the transfer of input data from the host system to the FPGA, executes the inference, and transfers the results back to the host system. Efficient data transfer protocols and mechanisms, such as direct memory access (DMA) and on-chip memory buffers, are utilized to minimize latency and maximize throughput. The communication interface design must be carefully engineered to minimize data transfer overhead and ensure seamless integration with the host system.

%% file: results.tex
\begin{table}[]
\centering
\caption{List of main InfoNEAT parameters set before the start of the evolution process.}\label{tab:infoneat}
\begin{tabular}{|c|c|}
\hline
\textbf{InfoNEAT Parameter}                                  & \textbf{Value}     \\ \hline
Population size (number of NNs per generation) & 16                 \\ \hline
Batch size                                          & 500                \\ \hline
Number of initial hidden nodes                      & 1                  \\ \hline
Maximum number of generations                       & 30                 \\ \hline
Variation of weights and biases ($\sigma$)             & 0.155              \\ \hline
Min., max., mean of weights and biases              & -3$\sigma$, 3$\sigma$, 0 \\ \hline
\end{tabular}
\end{table}
\begin{figure}[t]
\centerline{\includegraphics[width=0.46\textwidth,height=0.47\textwidth,keepaspectratio]{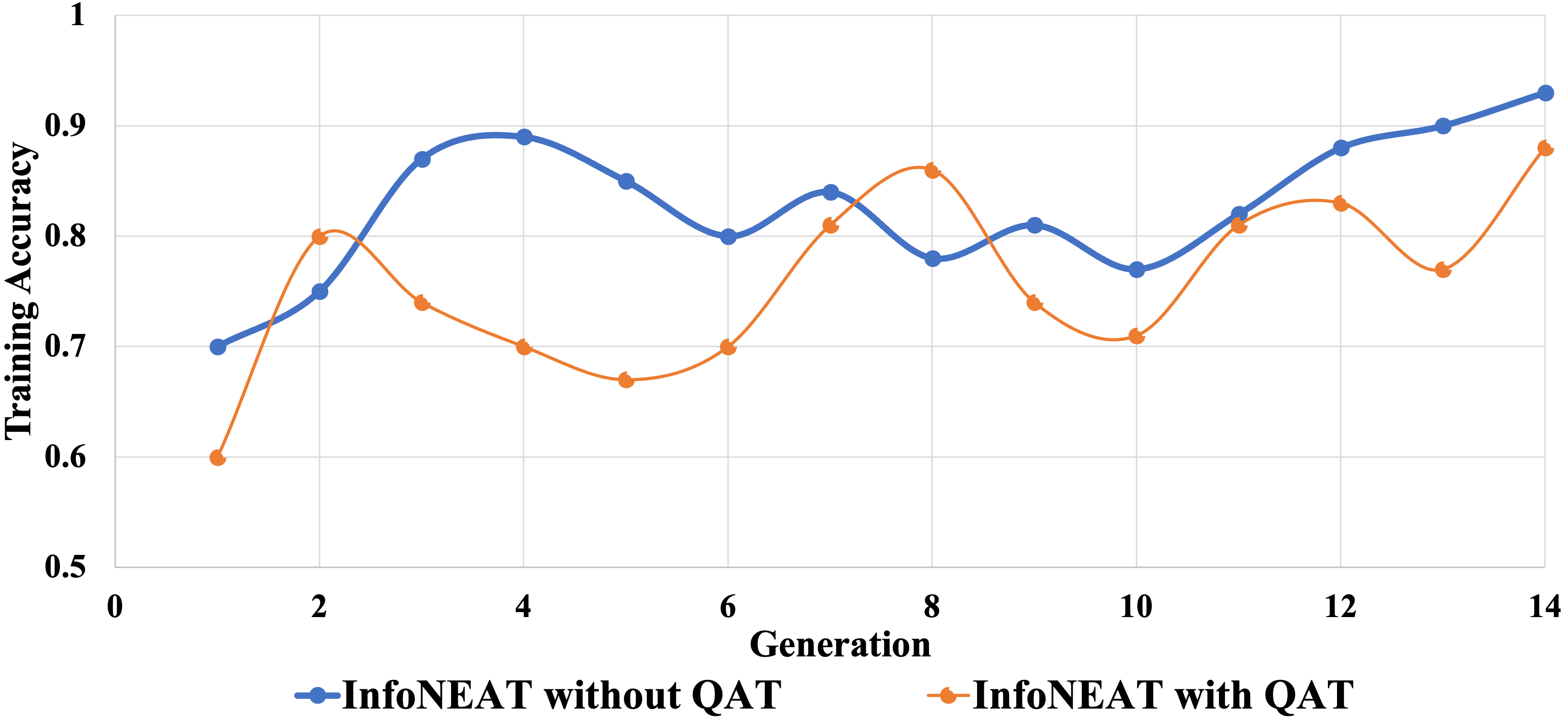}}
\caption[Learning curve of the InfoNEAT model against the intrusion detection dataset]{Learning curve of the InfoNEAT model with and without quantization across different generations during training.}
\label{fig:learning_curve}
\end{figure}
\section{Results and Analysis}\label{sec:results}
\subsection{Dataset}\label{sec:dataset}
We use the UNSW-NB15 dataset~\cite{moustafa2015unsw} created by the Australian Center for cybersecurity, which contains a large amount of data across various different attack modalities. This dataset contains data for normal network data and data that has been attacked using nine different methodologies -- fuzzers, analysis (corresponding to port access and scams, as named in the dataset), backdoors, exploits, reconnaissance, DoS, generic, shellcode, and worms. Each data row, representing a traffic flow, contains $49$~different features comprised of source and destination IP addresses, port numbers, transaction protocols, packet counts, jitters, etc. All these features can then be binarized as they are formatted differently, which then leads to a total of $593$~bits of features for the input~\cite{umuroglu2020logicnets}. 
We do not use all the $593$~bits but rather use $64$~header bytes of up to four consecutive packets in the same flow, as described in~\cite{le2022feature}. This leads to a total of $256$~bytes or features for the input which we use without any pre-processing such as normalization or standardization. We then create a binarized dataset with two labels -- attack and normal. Furthermore, we ensure that the training dataset contains an equal number of ``normal'' and ``attack'' samples. For this purpose, we take an equal number of ``attack'' data ($n_a$) from all nine different attack labels. Then, we take $9n_a$ number of ``normal" data to create this balanced binarized dataset.
\subsection{Experimental Setup}\label{sec:setup}
\begin{figure}[t]
\centerline{\includegraphics[width=0.4\textwidth,height=0.4\textwidth,keepaspectratio]{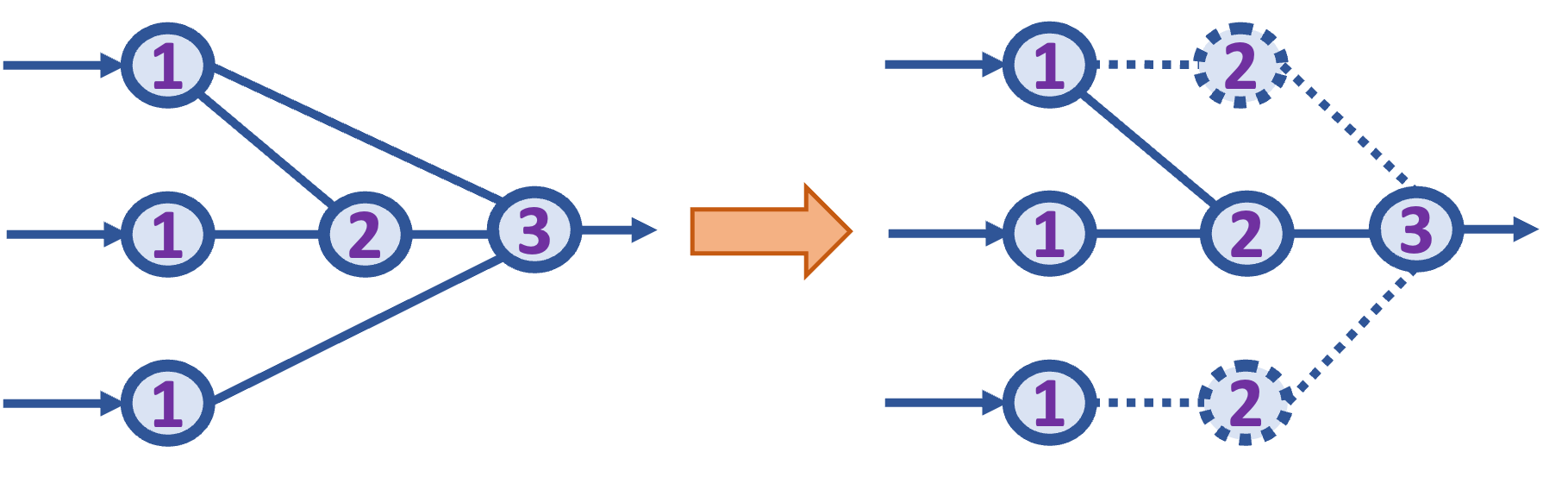}}
\caption[Modification of the q-InfoNEAT model for FPGA deployment]{Modification of the q-InfoNEAT model for FPGA deployment. The network on the left is a typical InfoNEAT model with its irregular architecture (such as a direct connection from layer 1 to layer 3). The network on the right is the modified model with an insertion of dummy nodes and connections (represented by dotted shapes) such that the resultant model resembles an MLP network. The number inside the shapes shows their corresponding layer number.}
\label{fig:dummy_mlp}
\end{figure}

As noted before, we taken advantage of the InfoNEAT's available code~\cite{Acharya_InfoNEAT} to further meet the needs for an IDS running on a hardware platform. 
We also use the QAT method~(Section~\ref{sec:method_qat}) within InfoNEAT to train quantized models with $2$-bit quantization. The code for LQ-Nets is available in~\cite{lq_net}. For our purpose, we first convert this tensorflow-based to a pytorch-based code. Then, we adapt the pytorch-based LQ-Net code within our InfoNEAT framework to design q-InfoNEAT. 
More specifically, we use the design methodology described in Section~\ref{sec:infoneat} to evolve and train InfoNEAT models using the dataset described in Section~\ref{sec:dataset}. Note that we do not perform any pre-processing on the features; rather we use the raw data directly. The main InfoNEAT parameters and their corresponding values are provided in Table~\ref{tab:infoneat}. The results and comparison against the state-of-the-art techniques will be explained in the upcoming subsections. The training and testing procedure implemented is as follows:
\begin{enumerate}[leftmargin=*,nosep]
    \item We first binarize the UNSW-NB15 dataset such that the dataset has two classes: ``normal'' and ``attack'' as described in Section~\ref{sec:dataset}. We then perform $k$-fold cross-validation analysis across different values of $k$ (ranging from $3$ to $10$) to figure out the optimal number of $k$-folds. Using this value of $k$, we split the dataset into balanced training and testing datasets.
    \item Next, we train $k$~different models using the training datasets created in the previous step. We first use InfoNEAT to evolve ``un-quantized'' models using the parameters shown above. Note that we use a batch of data per generation during training. Our analysis shows that a batch size of $500$ with an equal number of data per class provides the best result. Then, we use QAT within InfoNEAT to train quantized models. We use \textbf{accuracy} as our fitness metric to train and subsequently validate the models. Other ML metrics such as \textbf{F1-score} are also considered.
    \item Using the test dataset, we then evaluate our trained models using either accuracy or F1-score metrics.
    \item Finally, we implement the quantized model in hardware using FINN~\cite{blott2018finn, finn}, a framework developed by Xilinx which generates high-level synthesis~(HLS) code of quantized neural networks that can be accelerated using an FPGA. Because of the irregular structure of the model developed by q-InfoNEAT, the systolic arrays used by FINN for dense matrix-matrix multiplication cannot be directly used. Thus, we insert dummy nodes within the q-InfoNEAT network such that it mimics a regular MLP network~\cite{kao2021e3} as shown in Fig.~\ref{fig:dummy_mlp}. This method is inefficient compared to using an accelerator that can work with irregular NEAT networks. However, we use this method solely for comparison purposes and keep the efficient design of an accelerator part a focus for our future work. 
\end{enumerate}
\subsection{Results}
\subsubsection{Impact of using imbalanced dataset}
\begin{figure}[t]
\centerline{\includegraphics[width=0.47\textwidth,height=0.4\textwidth,keepaspectratio]{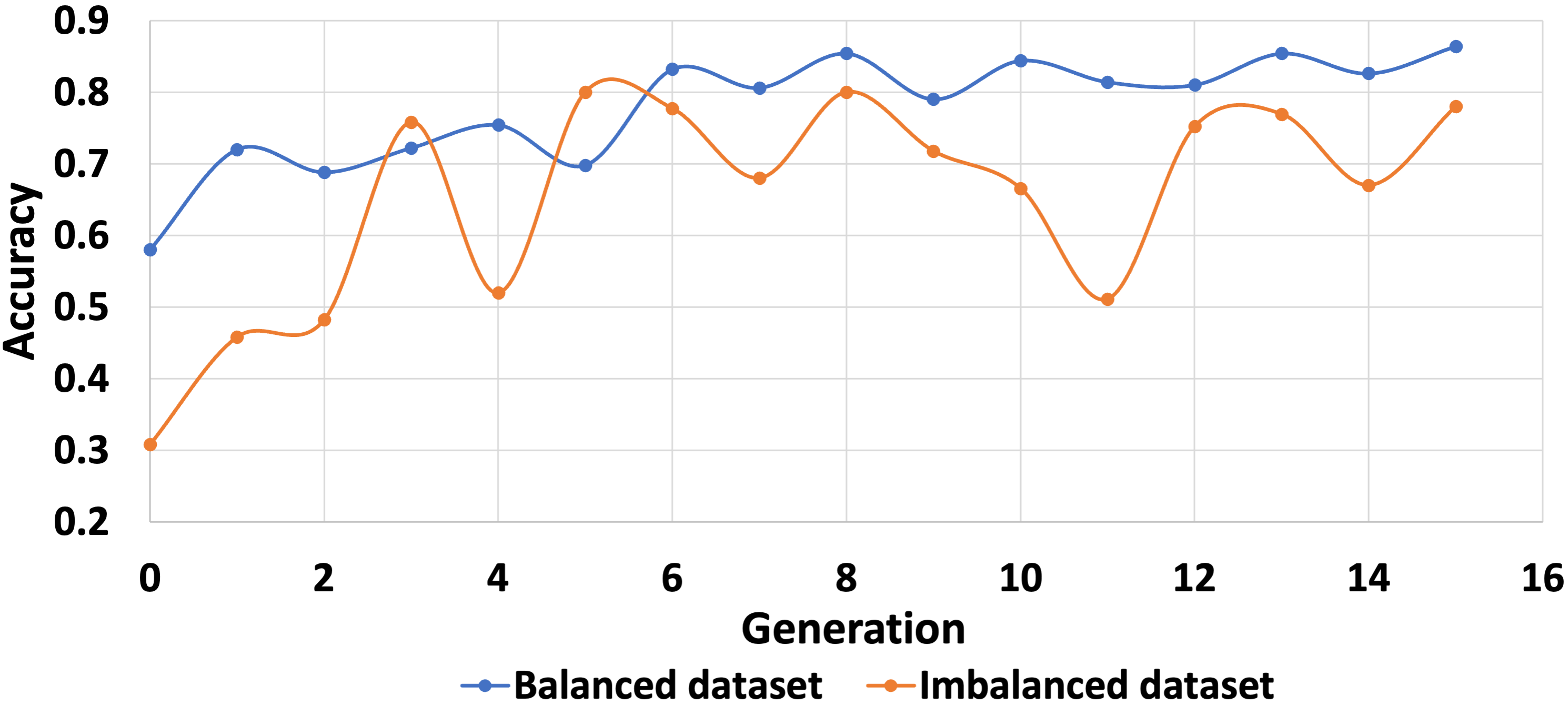}}
\caption{Comparison of training  q-InfoNEAT model on balanced versus imbalanced UNSW-NB15 dataset.\vspace{15pt}}
\label{fig:results_balanced_imbalanced}
\end{figure}

As discussed in Section~\ref{sec:dataset}, we train q-InfoNEAT on a balanced dataset. As a result, the training accuracy of q-InfoNEAT has low bias across different labels and low variation across different generations as shown in Fig.~\ref{fig:results_balanced_imbalanced}. Furthermore, we checked the robustness of these trained models by testing them against different day network data. The testing accuracy dropped from $0.86$ to $0.79$ when using an imbalanced dataset. Note that this check was performed using the complete dataset rather than different folds of cross-validation.
\subsubsection{Sensitivity to different $k$~folds of validation}
We also perform the sensitivity analysis using different $k$~folds of validation on the balanced UNSW-NB15 dataset. This is to figure out the optimal value for $k$ as well as to see how the algorithm behaves against different folds of the dataset. This in turn helps in providing a comprehensive assessment of the model's performance and robustness. Furthermore, it helps reduce the bias that might be introduced by using a single train-test split. From this analysis shown in Table~\ref{tab:sensitivity_analysis}, we can see that the average accuracy values do not differ a lot between different $k$~folds. Furthermore, the optimal value of $k$ is $4$ solely based on its corresponding highest average validation accuracy. 

\begin{table}[]
\small
\centering
\caption{Cross-validation results for different k-folds}
\label{tab:sensitivity_analysis}
\begin{tabular}{c|ccc|}
\cline{2-4}
\multicolumn{1}{l|}{} & \multicolumn{3}{c|}{\textbf{Accuracy}} \\ \hline
\multicolumn{1}{|c|}{\textbf{k-folds}} & \multicolumn{1}{c|}{\textbf{Minimum}} & \multicolumn{1}{c|}{\textbf{Maximum}} & \textbf{Average} \\ \hline
\multicolumn{1}{|c|}{3} & \multicolumn{1}{c|}{0.832} & \multicolumn{1}{c|}{0.909} & 0.877 \\ \hline
\multicolumn{1}{|c|}{4} & \multicolumn{1}{c|}{0.852} & \multicolumn{1}{c|}{0.947} & 0.884 \\ \hline
\multicolumn{1}{|c|}{5} & \multicolumn{1}{c|}{0.753} & \multicolumn{1}{c|}{0.894} & 0.857 \\ \hline
\multicolumn{1}{|c|}{6} & \multicolumn{1}{c|}{0.802} & \multicolumn{1}{c|}{0.932} & 0.880 \\ \hline
\multicolumn{1}{|c|}{7} & \multicolumn{1}{c|}{0.824} & \multicolumn{1}{c|}{0.937} & 0.873 \\ \hline
\multicolumn{1}{|c|}{8} & \multicolumn{1}{c|}{0.823} & \multicolumn{1}{c|}{0.940} & 0.868 \\ \hline
\multicolumn{1}{|c|}{9} & \multicolumn{1}{c|}{0.798} & \multicolumn{1}{c|}{0.936} & 0.879 \\ \hline
\multicolumn{1}{|c|}{10} & \multicolumn{1}{c|}{0.778} & \multicolumn{1}{c|}{0.894} & 0.848 \\ \hline
\end{tabular}%
\end{table}

\subsubsection{Learning curve}
Fig.~\ref{fig:learning_curve} shows the evolution of the q-InfoNEAT model across different generations. 
Based on Fig.~\ref{fig:learning_curve}, we can see that the accuracy values are slightly worse for the model with quantization compared to that without quantization. Furthermore, training InfoNEAT with quantization seems to introduce higher variance in performance~(or accuracy) across different generations.


\begin{table}[t]
\centering
\scriptsize
\caption{Comparison of the network size, accuracy (Acc.), number of LUTs, latency, and throughput (Thruput) used for q-InfoNEAT model against state-of-the-art techniques. The number of classes used in all these works is $2$. IFS, PPS, FPS refer to inferences per second, packets per second, and features per second, respectively.}
\begin{tabular}{|c|c|c|c|c|c|}
\hline
\textbf{Type of NN} & \textbf{Size} & \textbf{Acc.} & \textbf{LUTs} & \textbf{Latency} & \textbf{Thruput} \\ \hline
BNN~\cite{murovivc2021genetically} & 593$\times$64$\times$1 & 0.92 & 26,879 & 19 ns & \makecell{256G\\bps} \\ \hline
LogicNets~\cite{umuroglu2020logicnets} & \begin{tabular}[c]{@{}c@{}}593$\times$256$\times$\\128\ $\times$128$\times$1\end{tabular} & 0.91 & 15,949 & 10.5 ns & - \\ \hline
BNN~\cite{vrevca2021detecting} & \begin{tabular}[c]{@{}c@{}}42$\times$80$\times$\\200$\times$1\end{tabular}  & 0.82 & 26,556 & 91 ns & \begin{tabular}[c]{@{}c@{}}10.98M\\IFS\end{tabular} \\ \hline
1D CNN~\cite{le2022feature} & \begin{tabular}[c]{@{}c@{}}256$\times$4096$\times$\\2048$\times$ 4096$\times$\\1536$\times$96$\times$13\end{tabular} & 0.987 & 22,893 & 1.78 ms & \makecell{2,428\\PPS} \\ \hline
2D CNN~\cite{le2022feature} & \begin{tabular}[c]{@{}c@{}}256$\times$4096$\times$\\2048$\times$6144\\$\times$4096$\times$256\\$\times$13\end{tabular} & 0.998 & 59,679 & 85.26 us & \makecell{86,672\\PPS} \\ \hline
DWS CNN~\cite{le2022feature} & \begin{tabular}[c]{@{}c@{}}256 $\times$1024 $\times$\\2048$\times$512$\times$\\1024$\times$64\\$\times$13\end{tabular} & 0.998 & 47,297 & 6.68 us & \makecell{1.9M\\PPS} \\ \hline
q-InfoNEAT & \begin{tabular}[c]{@{}c@{}}256$\times$4$\times$1\end{tabular} & 0.947 & 6,943 & 113 ns & \makecell{97.1M\\FPS} \\ \hline
\end{tabular}
\label{tab:results}
\end{table}

\subsubsection{Comparison against state-of-the-art}
Table~\ref{tab:results} shows and compares the results of different NNs against the q-InfoNEAT model. These include state-of-the-art BNNs~\cite{murovivc2021genetically, vrevca2021detecting}, LogicNets (a type of quantized deep NN)~\cite{umuroglu2020logicnets}, and CNNs~\cite{le2022feature}.The table shows the size of the network which consists of the number of nodes in each layer. For the CNNs, we instead give the number of elements in each feature map. Based on this result, we can see that the q-InfoNEAT models have a much smaller architectural size compared to other NNs. This size is at most thousands times smaller and at least $16\times$ smaller compared to the other state-of-the-art networks. Similarly, the number of layers used to train InfoNEAT models is also comparatively less. 
However, this comes at a cost of performance which is only slightly degraded when comparing q-InfoNEAT to other networks. 
It is worth mentioning that the accuracy reported in Table~\ref{tab:results} is the maximum accuracy over the folds (average over folds is 0.87). 
We reported the maximum accuracy similar to the state-of-the-art, which has not considered $k$-fold cross-validation, making a direct comparison between average values impossible. 

Furthermore, we also compare the amount of hardware resources used by the NN once it is synthesized and deployed on an FPGA using FINN. This synthesis is performed at a clock frequency of $100$MHz using the ZCU104 FPGA platform. The resource consumption is then calculated based on the implementation using Xilinx Vivado 2023.1 and are shown as lookup tables~(LUTs) in Table~\ref{tab:results}. 
A LUT is a small memory unit capable of storing and retrieving data based on input combinations. 
FPGAs consist of an array of configurable logic blocks~(CLBs) which essentially consist of LUTs that are designed to represent any arbitrary combinational logic by programming the desired truth table into its memory. Thus, the number of LUTs utilized by the model gives a good idea of how much hardware resource is used by the NN. 
Based on Table~\ref{tab:results}, we can see that the q-InfoNEAT model after quantization uses the fewest number of LUTs. The number of LUTs utilized is between $2.3\times$~and~$8.5\times$ smaller compared to other proposed techniques. 

%% file: conclusion.tex
\section{Conclusion}\label{sec:conclusion}
Deploying ML-based IDSs on hardware devices poses a significant challenge due to their limitations in computational resources, power consumption, and network connectivity. In this paper, we introduced a design methodology that automatically trains and evolves a quantized neural network (NN) model that is a thousand times smaller than the current state-of-the-art NN. Furthermore, this network, when deployed on hardware, has lower latency, higher throughput, and subsequently utilizes up to $8.5\times$ fewer LUTs while achieving comparable performance when compared to the state-of-the-art techniques.